\pdfoutput=1 
\documentclass[aps,prl,twocolumn,showpacs]{revtex4-1}
\usepackage{graphicx}
\setcitestyle{super}

\usepackage{graphicx}
\usepackage{color}
\usepackage{bm}
\usepackage{braket}
\usepackage{amsmath}
\usepackage{natbib}

\begin{document}


\title{
Exotic heavy fermion superconductivity in atomically thin CeCoIn$_5$ films}
\author{L. Peng$^{1}$, M. Naritsuka$^{1,2}$,  S. Akutagawa$^{1}$,  S. Suetsugu$^{1}$,  M. Haze$^{3}$, Y. Kasahara$^{1}$, T. Terashima$^{1}$, R. Peters$^1$,  Y. Matsuda$^{1}$, and T. Asaba$^{1*}$}

\affiliation{
$^1$Department of Physics, Kyoto University, Kyoto 606-8502, Japan.\\
$^2$School of Physics and Astronomy, University of St Andrews, North Haugh, St Andrews, Fife KY16 9SS, UK.\\
$^3$Institute for Solid State Physics, University of Tokyo, Kashiwanoha 5-1-5, Kashiwa, Chiba 277-8581,  Japan.\\
}
\date{\today}
\begin{abstract}

We report an {\it in-situ} scanning tunneling microscopy study of atomically thin films of CeCoIn$_5$, a $d$-wave heavy-fermion superconductor. Both hybridization and superconducting gaps are observed even in monolayer CeCoIn$_5$, providing direct evidence of  superconductivity of heavy quasiparticles mediated by purely two-dimensional bosonic excitations.  In these atomically thin films,  $T_c$  is suppressed to nearly half of the bulk, but is similar to CeCoIn$_5$/YbCoIn$_5$ superlattices containing CeCoIn$_5$ layers with the same thickness as the  thin films.  Remarkably, the out-of-plane upper critical field $\mu_0H_{c2\perp}$ at zero temperature is largely enhanced from those of bulk and superlattices.   The enhanced $H_{c2\perp}$ well exceeds the Pauli and bulk orbital limits, suggesting the possible emergence of unusual superconductivity with parity mixing caused by the inversion symmetry breaking.
\end{abstract}

\maketitle

 The superconducting properties of thin films can dramatically deviate  from their bulk behavior when the film thickness is reduced to the atomic scale~\cite{qing2012interface,xi2016ising,liu2018interface,xing2017ising}.   In sharp contrast to  most of bulk superconductors,  atomically thin films naturally break inversion symmetry.   In the presence of strong spin-orbit interaction (SOI), these non-centrosymmetric superconductors  have aroused significant interest because of their  non-trivial superconducting states, including parity-mixing pairing states~\cite{gor2001superconducting,frigeri2004superconductivity}, non-reciprocal superconducting properties~\cite{yuan2021supercurrent,daido2022intrinsic} and  topological superconductivity~\cite{qi2011topological,tanaka2009theory,sato2010existence,yoshida2016topological,daido2017majorana}.    Of particular interest is the atomically thin two-dimensional (2D) systems of  strongly correlated superconductors with non-$s$-wave pairing.  It has been pointed out that the effect of the inversion symmetry breaking in such superconductors may be more pronounced than that in weakly correlated conventional superconductors with $s$-wave symmetry~\cite{fujimoto2007fermi,fujimoto2007electron,maruyama2015electron}.

CeCoIn$_5$ is a heavy-fermion compound that hosts a wide range of fascinating unconventional superconducting properties   ($T_c$ =2.3\,K),  such as  extremely strong coupling $d_{x^2-y^2}$ superconductivity~\cite{petrovic2001heavy,PhysRevLett.100.087001,an2010sign,allan2013imaging,zhou2013visualizing} and a strong Pauli paramagnetic pair breaking effect~\cite{PhysRevLett.87.057002,bianchi2002first,bianchi2008superconducting}.  Despite its layered structure  shown in Fig.\,1(a), the corrugated Fermi surface~\cite{hall2001electronic,hall2001fermi,settai2001quasi,settai2007recent} and 3D-like antiferromagnetic spin-fluctuations characterized by the wavenumber {\boldmath $q$}=(0.45, 0.45, 0.5)~\cite{stock2008spin,kenzelmann2008coupled,raymond2015ising} suggest that the electronic and magnetic properties are rather 3D than 2D. Therefore, clarifying how the superconducting properties change in the pure 2D limit, including the fate of superconductivity, is a key to understanding the superconducting mechanism of CeCoIn$_5$. To tackle this issue, we fabricated Kondo superlattices~\cite{shishido2010tuning} consisting of alternating layers of CeCoIn$_5$ and conventional nonmagnetic metal YbCoIn$_5$~\cite{mizukami2011extremely,shimozawa2016kondo,naritsuka2017emergent}. However, the goal to understand the properties of a single superconducting CeCoIn$_5$ layer inside a superlattice was not achieved due to the lack of zero resistance observed in mono- and bilayer CeCoIn$_5$ superlattices, and it remains open whether CeCoIn$_5$ is superconducting in the 2D limit. Therefore, it has been strongly desired to fabricate atomically thin films of CeCoIn$_5$.  Moreover, as the SOI is generally strong in Ce-compounds, the introduction of the inversion symmetry breaking is expected to make the systems a fertile ground for observing exotic properties~\cite{yuan2014probing,daido2017majorana,yoshida2015topological}.  

Here,  by using a state-of-the-art molecular beam epitaxy (MBE) technique~\cite{shishido2010tuning},  atomically thin films of CeCoIn$_5$  were epitaxially grown on the surface of non-magnetic metal films of YbCoIn$_5$ (80 nm-thick), which were grown on the (001) surface of the substrate MgF$_2$ (Fig.\,1(b)).  The films were transferred {\it in-situ} to the low-temperature scanning tunneling microscopy and spectroscopy (STM/STS)  head after deposition.   The  low-temperature STM/STS techniques reveal the appearance of  hybridization and superconducting gaps even in monolayer CeCoIn$_5$, indicating the heavy fermion superconductivity mediated by purely 2D bosonic excitations.     The superconducting properties of the heavy-fermion thin film are dramatically different from CeCoIn$_5$/YbCoIn$_5$ superlattices and bulk CeCoIn$_5$, suggesting that inversion symmetry breaking largely influences the superconducting properties.

The typical STM topographic image of a CeCoIn$_5$ atomically thin film  is displayed in Fig.\,1(c).  Shown in Fig.\,1(d) is the cross-sectional profile along the red arrow in Fig.\,1(c). A double step structure is clearly observed.   The second step height is 0.75\,nm, which coincides with the $c$-axis lattice constant of bulk CeCoIn$_5$.   The inset of Fig.\,1(c) depicts the atomic resolution STM topograph taken on the terrace C. The spatially periodic bright spots forming the square lattice represent the In atoms in the Ce-In plane.   We note that the local electronic structure is measured mainly through the $p_z$ orbital of the In atoms which is well extended perpendicular to the surface.  In fact, the distance between these spots determined by this image is 0.45\,nm, which coincides with the In-In distance in this plane.   In Fig.\,1(d), the first step located in the middle of the Ce-In planes corresponds to the Co layer, which is consistent with the previous report on bulk CeCoIn$_5$~\cite{zhou2013visualizing} (see Fig.\,S3 for details~\cite{SI}).   

\begin{figure}[t]
	\centering
	\includegraphics[width=\linewidth]{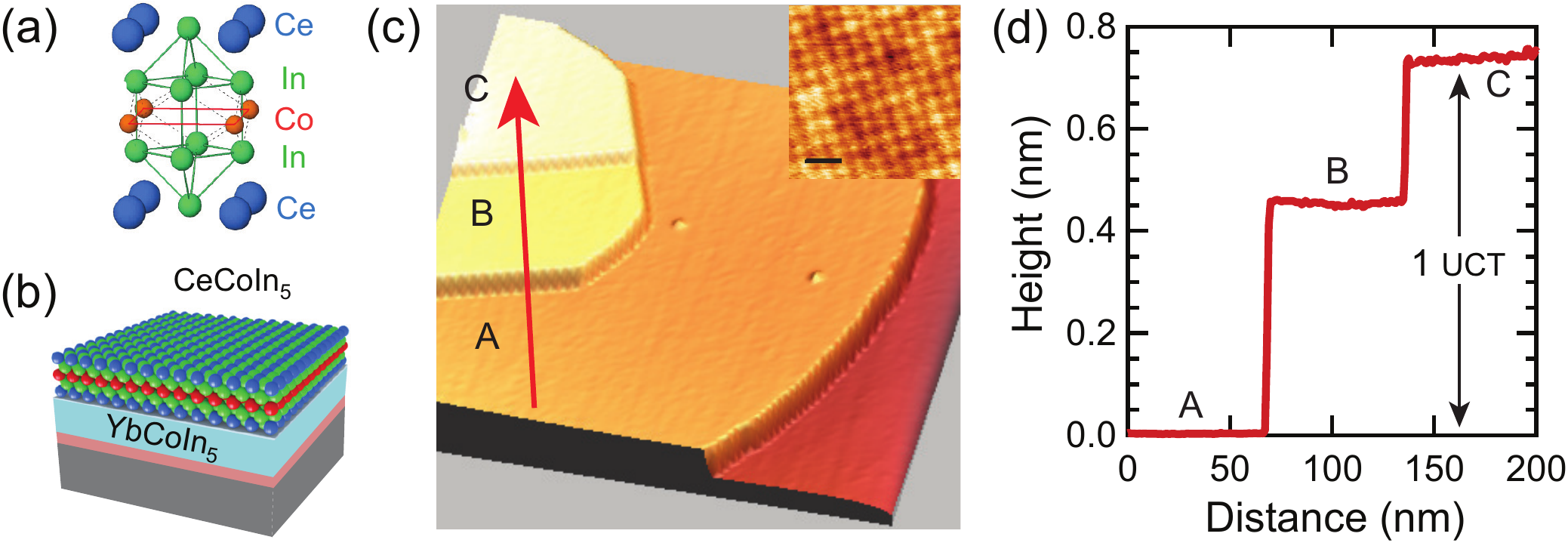}
	\caption{\label{Fig:basic}  (a) Crystal structure of CeCoIn$_5$. (b) Atomically thin film of  CeCoIn$_5$ epitaxially grown on nonmagnetic conventional metal YbCoIn$_5$.  YbCoIn$_5$ is epitaxially grown on a MgF$_2$ substrate (gray layer) and CeRhIn$_5$ (pink layer)is used as a buffer layer between YbCoIn$_5$ and MgF$_2$.  (c)  3D STM topographic image (sample bias voltage $V_s = 1.0$ V, tunneling current $I_t = 50$ pA.) of a bilayer CeCoIn$_5$  thin film (230\,nm $\times$\,230 nm). The data is taken at $T$ =\,1.6\,K.  Brown region corresponds to the YbCoIn$_5$ layer.   Terraces A and C represent the first and second Ce-In layer, respectively.  Terrace B represents the Co layer of the second CeCoIn$_5$.     (inset) Atomic resolution STM image ($V_s = $100 mV, $I_t = $50 pA) taken on the terrace C, where white bright spots forming the square lattice represent the In atoms in the Ce-In plane.  Scale bar: $1$ nm. (d) The cross-sectional height profile along the red arrow in Fig.\,1(c).  A double-step structure is clearly observed.   The second step height is 0.75\,nm, which coincides with the $c$-axis lattice constant of CeCoIn$_5$. }
\end{figure}
Figure\,2(a) displays the high signal-to-noise ratio differential tunnelling conductance d$I$/d$V$ spectra, which are proportional to the local density of states (LDOS), on the Ce-In layers of CeCoIn$_5$ ranging from one to six unit-cell-thickness (see Fig.\,S4 for details~\cite{SI}) at $T$=1.6\,K.    For the comparison,  spectra at the Co layer of CeCoIn$_5$, the Yb-In surface of YbCoIn$_5$ and the Ce-In surface of CeCoIn$_5$ thick film with a thickness of order of 100\,nm are also shown.   A clear gap structure centered at around 7\,mV and 0\,mV can be seen on Ce-In and Co layers, respectively, while such a gap is absent on YbCoIn$_5$.    The reduction of the LDOS  around the Fermi energy $E_F$ on the Ce-In layer arises from the formation of the hybridization gap $\Delta_{H}\sim$20\,meV.  As the temperature is lowered below the Kondo temperature $T_K$, the 4$f$-electrons hybridize with the conduction electrons via the Kondo effect, opening a gap, and collective coherent screening takes place.  The Kondo temperature $T_K$ is about 40\,K for bulk CeCoIn$_5$~\cite{Nakajima2007}.  A hybridization gap with a slightly different shape is also observed on the Co layer, which will be discussed later. The magnitude of $\Delta_H$ of the Ce-In layers in the atomically thin films is comparable or slightly larger than that in single crystals~\cite{allan2013imaging,zhou2013visualizing,gyenis2018visualizing} and  thick films~\cite{Haze2018}.   It should be stressed that the hybridization gap provides direct evidence for the formation of the heavy quasiparticle bands.  Furthermore, its magnitude represents the strength of the hybridization between $f$-  and conduction electrons and is directly related to the effective electron mass.  Therefore, a similar value of $\Delta_H$ in atomic thin films implies the formation of heavy quasiparticles with similar mass as in the bulk.

\begin{figure}[t]
	\includegraphics[width=\linewidth]{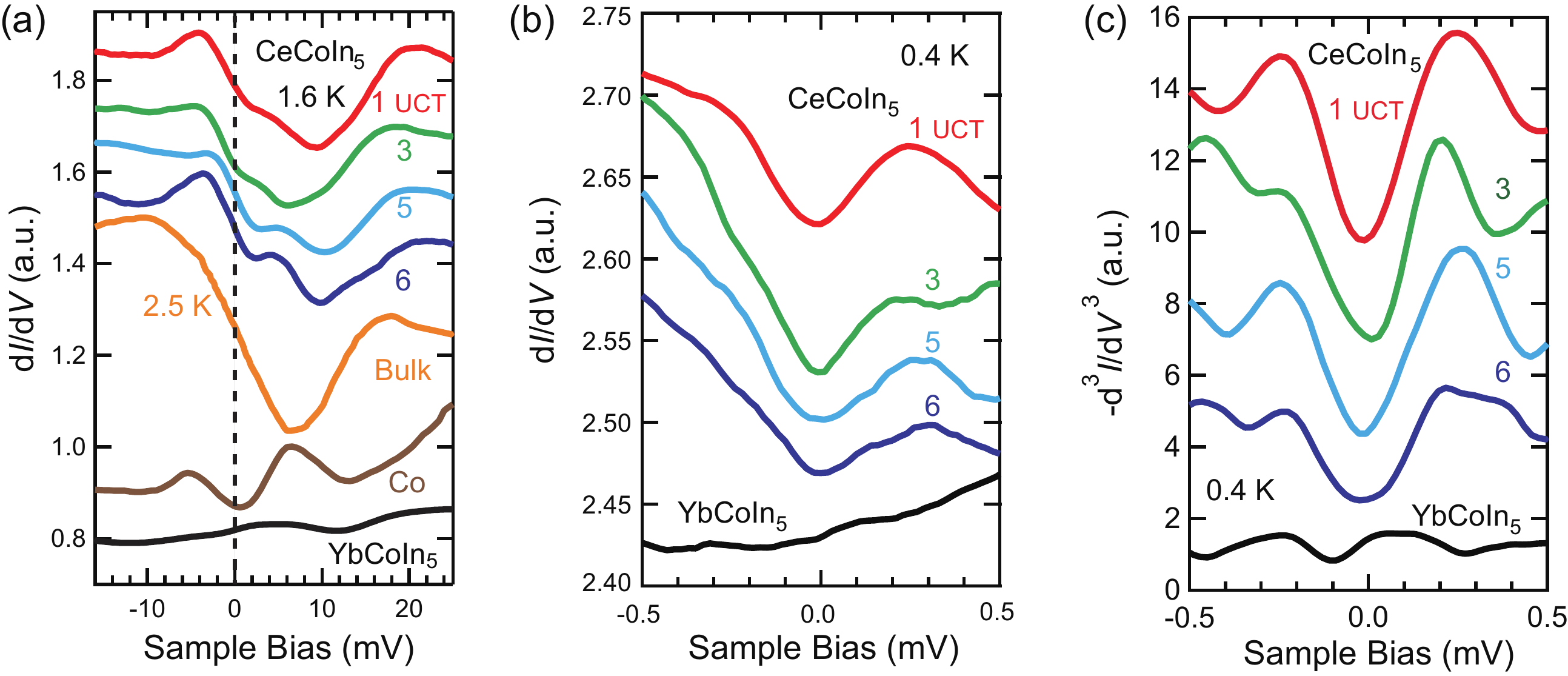}
	\caption{\label{fig:layers} 
	 (a) Differential tunnelling conductance d$I$/d$V$ spectra on the Ce-In layers of CeCoIn$_5$  ranging from one to six unit-cell-thickness (UCT) at $T$=1.6\,K.    For the comparison,  spectra at the Co layer at terrace B in Fig.\,1(c) and YbCoIn$_5$ and bulk CeCoIn$_5$ are also shown.  A clear gap structure can be seen on both Ce-In and Co layers, while it is absent on YbCoIn$_5$. Tunnelling parameters: $V_s =  30$ mV, $I_t = 100$ pA, $V_{mod}=300 $ $\mu$V.
		(b)  d$I$/d$V$ spectra  of atomically thin films of CeCoIn$_5$ in the low bias regime at $T$=0.4\,K, along with the data of YbCoIn$_5$.    A clear superconducting gap structure is observed at $E_F$  even in monolayer CeCoIn$_5$, while no gap is observed on YbCoIn$_5$. Tunnelling parameter:  $V_s=2$ mV, $I_t = 200$ pA, $V_{mod}=50 $ $\mu$V.  Spectra are vertically shifted for clarity. (c) Sample bias dependence  of the second derivative of the tunnel conductance d$^3I$/d$V^3$ at different thicknesses.  Derivatives are taken after smoothing the data.
	}
	
\end{figure}
\begin{figure*}[t]
	\includegraphics[width=1\linewidth]{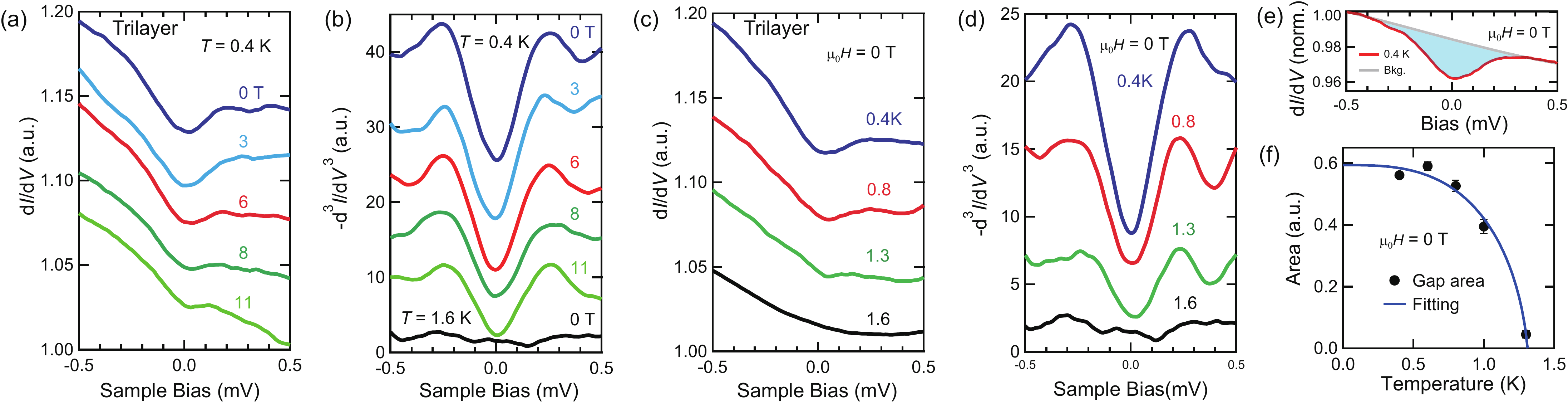}
	\caption{\label{fig:spectra} 
		 (a) Magnetic filed dependence of the d$I$/d$V$ spectra  at $T = $\,0.4\,K for {\boldmath $H$}$\parallel c$.  (b) Sample bias dependence  of the second derivative of the tunnel conductance d$^3I$/d$V^3$ at different magnetic fields.   The superconducting gap survives even at $\mu_0H =$\,11\,T. (c) Temperature dependence of  tunnelling conductance d$I$/d$V$ spectra  in zero magnetic field.  (d) Sample bias dependence  of the second derivative of the tunnnel conductance d$^3I$/d$V^3$ at different temperatures.  For (b) and (d), derivatives are taken after smoothing the data. (e) d$I$/d$V$ spectra at 0.4\,K,  normalized by that at 1.6\,K.   Light blue region corresponds to superconducting gap area.    The gray line represents the background obtained by cubic fitting in the high bias regime. (f) Temperature dependence of the superconducting gap area (see SI for details).   From the fitting with $S(T) \propto \sqrt{1-(T/T_c)^3}$,  $T_c\sim$ 1.32\,K is obtained.  The spectra are vertically shifted for clarity.  Tunnelling parameters: $V_s=2$ mV, $I_t=50$ pA, $V_{mod}=30$ $ \mu $V.
	}
\end{figure*}
Despite a similar hybridization gap between atomically thin films and bulk CeCoIn$_5$,  there are several  differences.  As shown in Fig.\,2(a), d$I$/d$V$ spectra show a shoulder structure at $\sim$2\,mV.   We note that such a structure is absent in bulk and thick films.  Moreover, the hybridization gap at the Co layer is much larger than that reported in single crystals~\cite{zhou2013visualizing}. This implies the stronger coupling between 4$f$ and Co 3$d$ conduction electrons in atomically thin CeCoIn$_5$.   When approaching the 2D limit, the DOS naturally changes, resulting in the modified hybridization gap. The difference in the system's symmetry (from 3D to 2D) may also be responsible for crystal field in-gap states, but further theoretical studies are required to understand this behavior fully. d$I$/d$V$ spectra reveal that in atomically thin films of CeCoIn$_5$,  although the electronic structure appears to be modified from that of bulk,  a similar heavy quasiparticle band as that of bulk is formed. 

Figure\,2(b) depicts d$I$/d$V$ spectra of the smaller gap centered around $V$=0 at the atomically flat terraces of  one-, three-, five- and six- unit-cell-thick CeCoIn$_5$, along with the spectra of YbCoIn$_5$,  at low bias at $T$=0.4\,K.    A distinct gap structure is observed at $E_F$ on all the CeCoIn$_5$ surfaces, while it is absent on the YbCoIn$_5$ surface.   Figures\,3 (a) and (c) depict the magnetic field and temperature dependencies of $dI/dV$ spectra of trilayer CeCoIn$_5$, respectively.  All d$I$/d$V$ data have a non-symmetric background arising from the energy dependent density of states.  To extract the detailed gap structure from the data, we take the second derivative of d$I$/d$V$, eliminating the linear background.    As shown in Figs.\,2(c), 3(b) and 3(d),  -d$^3I$/d$^3V$ is nearly symmetric to the bias voltage, suggesting the symmetric gap structure.  We have also tried to first fit the data using Gaussian+polynomial, then taken the second derivative of the fitted curves (see SI). The particle-hole symmetry is also observed, indicating that our analysis is valid.

The smaller gap shown in Fig.\,2(b) is most likely to be associated with a superconducting origin because of the following reasons.  First,  as shown in Fig.\,3(a), the gap is suppressed by the magnetic field  applied perpendicular to the 2D plane (See Fig.\,S8~\cite{SI}). Second,  as shown in Figs.\,3(c) and (d), the gap decreases with increasing temperature and is fully suppressed at 1.6\,K. After normalizing by $T$=1.6\,K data and subtracting the cubic background, the gap area at each temperature is obtained as shown in Fig.\,3(e) (See Fig. S3 for details). To determine the onset temperature of the gap, we plot the area of the gap region $S$ as a function of temperature in Fig.\,3(f) and fit the data by using $S(T) \propto \sqrt{1-(T/T_c)^3}$, resulting in $T_c\sim$1.3\,K~\cite{dora2001thermodynamics}.  It has been theoretically pointed out that when the Fermi surface is large enough~\cite{cappelluti2007topological}, which is the case for CeCoIn$_5$,  the inversion symmetry breaking does not change $T_c$ significantly.   We then  fabricated and measured the transport properties of a superlattice with the global inversion symmetry, in which trilayer CeCoIn$_5$ is sandwiched by five layers of nonmagnetic metal YbCoIn$_5$, as shown in Fig.\,S10~\cite{SI}. A clear superconducting transition is observed, with almost the same onset $T_c$ $\approx$1.3\,K as the trilayer thin film measured by STM.  Third, as shown in Figs.\,2(c), 3(b) and 3(d), the small gap from atomically thin CeCoIn$_5$ films is particle-hole symmetric for all thicknesses, magnetic fields, and temperatures after eliminating the linear background. As this feature is universal under different conditions,  we exclude the possibility of an extrinsic origin. It should be stressed that electron-particle symmetry is a fundamental property of the superconductivity.   Fourth, the gap value $\Delta$ roughly estimated from the peak of -d$^3I$/d$V^3$ is $\sim$ 0.25\,meV at 0.4\,K in zero field  in atomically thin films.  The magnitude of 2$\Delta/k_B T_c\sim$4.5 is close to the bulk value. Finally, as shown in the SI, the small gap is observed consistently on the CeCoIn$_5$ monolayer,  while the absence of the gap on YbCoIn$_5$ layer is firmly confirmed by the high signal-to-noise ratio measurement in Fig.\,2(b), excluding the possibility of the tip origin or site dependence.

We  point out that spin-density-wave (SDW) or pseudogap formation as the origin of the smaller gap is highly unlikely for the following reasons.  Since the hybridization gap is strongly asymmetric with respect to the Fermi energy as shown in Fig.\,2(a), the band structure has no particle-hole symmetry. Then, in such an asymmetric band structure, the gap associated with the SDW formation should also be asymmetric.  Moreover, the size of the gap of monolayer 2$\Delta\sim$0.5\,meV is comparable to that of six layers. This also appears to be incompatible with the SDW scenario, because the gap size of SDW is expected to change when the system approaches the 2D limit.  Moreover,  the pseudogap, which usually does not have particle-hole symmetry, has been reported only on the Co layers~\cite{zhou2013visualizing,gyenis2018visualizing}, while the present measurements are performed on the Ce-In layers. 

Based on these results, we conclude  that atomically thin films of CeCoIn$_5$ exhibit superconductivity down to monolayer thickness.  We note that  d$I$/d$V$ remains large at $E_F$ even at 0.4\,K well below $T_c$, indicating a large residual quasiparticles weight within the superconducting condensate. Such a large residual DOS has also been reported in thick films and single crystals~\cite{Haze2018,zhou2013visualizing}, which appear to be attributed to the multiband nature of the superconductivity. In fact, in tetragonal FeSe$_{1-x}$S$_x$\cite{hanaguri2018two} and UTe$_2$~\cite{jiao2020chiral}, the zero-bias DOS is very high even at $T/T_c\sim 0.3$, at which we performed the STM measurements. In our case, the multiband effect is expected to be further enhanced by Rashba SOC and two-dimensionality. The former enhances the multigap effect through the splitting of the Fermi surfaces and the latter reduces the freedom in the {\boldmath $k$}-vector to form Cooper pairs, resulting in a higher portion of ungapped Fermi surface sheets in a wide temperature range in the superconducting state. We also point out that the coherence peak is smeared out by the same reasoning.

\begin{figure}[t]
	\includegraphics[width=0.8\linewidth]{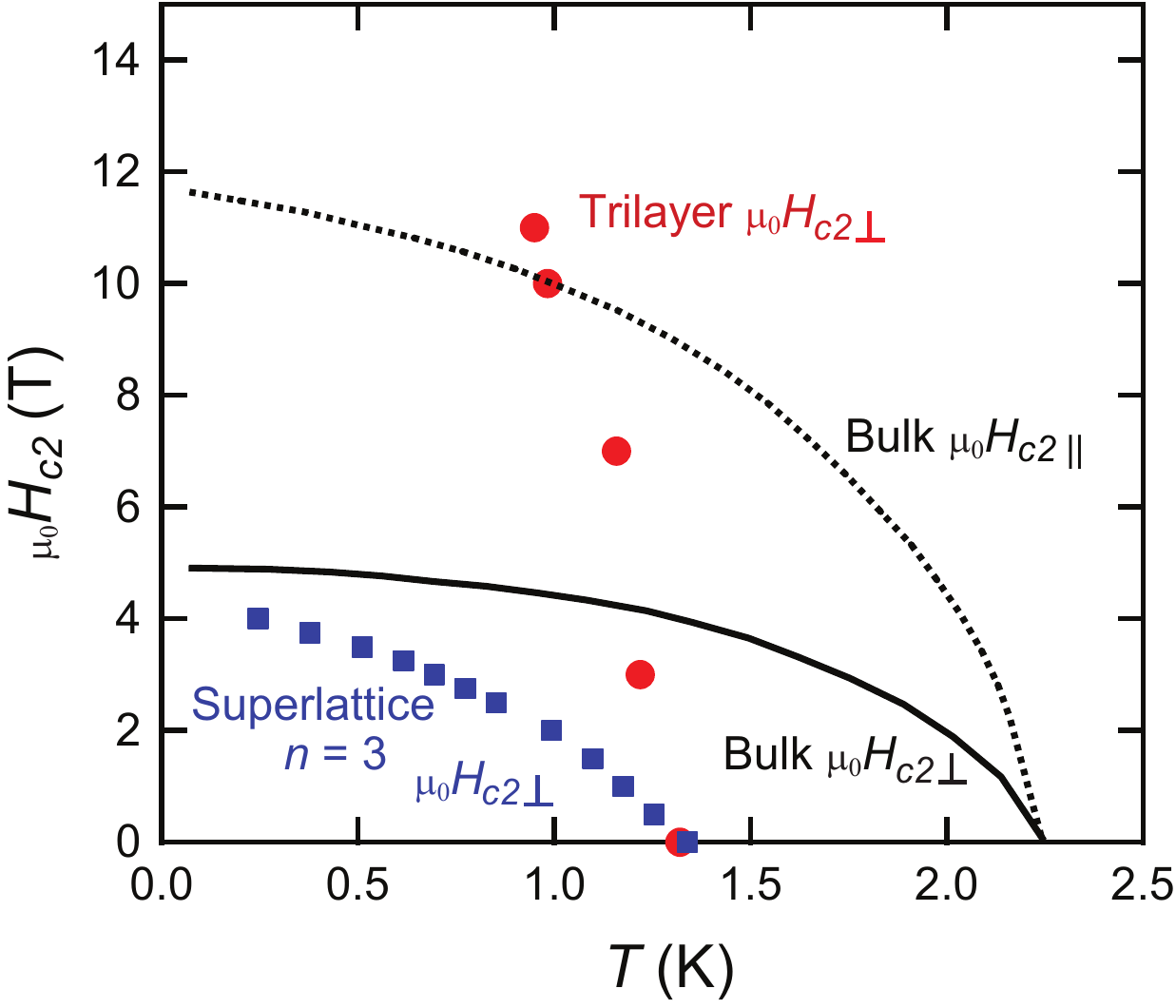}
	\caption{\label{fig:phase} 
		 Temperature dependence of out-of-plane upper critical field $H_{c2\perp}$ of trilayer CeCoIn$_5$ film determined by the tunnelling conductance (red circle).  For the comparison, $H_{c2\perp}$ for the bicolor CeCoIn$_5$($n$)/YbCoIn$_5$(5) superlattice with $n$ =3 (blue square) and  $H_{c2\perp}$ (solid line) and in-plane upper critical field $H_{c2\parallel}$ (dashed line) of the CeCoIn$_5$ single crystals are also shown. For the superlattice data, shown are the superconducting onset temperatures defined by $R(T_{onset})=0.9\,R_{normal}$ determined from transport measurements. 
	}
\end{figure}

Figure\,4 depicts the $T$-dependence of the out-of-plane upper critical field $H_{c2\perp}$ of trilayer CeCoIn$_5$, which is determined by measuring the $T$-dependence of the area of the superconducting gap region at each field (see Fig.\,S7 and Fig.\,S8 for details~\cite{SI}). For the comparison, the results from the bulk and the CeCoIn$_5$(3)/YbCoIn$_5$(5) superlattice, where superconductivity has been confirmed by transport measurements, are also shown~\cite{mizukami2011extremely}. What is most remarkable is that although $T_c$ at zero field is largely reduced from the bulk value, $H_{c2\perp}$ strikingly exceeds $H_{c2\perp}$ of the bulk and is even higher than the in-plane upper critical field $H_{c2\parallel}$.   In bulk CeCoIn$_5$, the superconductivity is limited by a strong Pauli paramagnetic effect for both field directions; $\mu_0H_{c2\perp}\approx \mu_0H_{Pauli\perp}\approx 5$\,T.  The orbital-limited upper critical field of bulk CeCoIn$_5$ obtained from the initial slope of the upper critical field at $T_c$, $H_{c2\perp}^{orb}=-0.73\,T_c(dH_{c2\perp}/dT)|_{T_c}$, is $\mu_0H_{c2\perp}^{orb}\approx 15$\,T.  Surprisingly, the orbital-limited upper critical field of trilayer CeCoIn$_5$ is estimated to be  $\mu_0H_{c2\perp}^{orb}\approx 30$\,T, nearly twice as large as the bulk.  Thus the present results demonstrate that not only the Pauli paramagnetic effect but also the orbital pair-breaking effect are strikingly suppressed in atomically thin films. Since such an enhancement was not observed in the superlattice with the same CeCoIn$_5$ thickness and similar $T_c$, this enhancement cannot be simply attributed to the dimensionality effect. Therefore, the enhancement of the upper critical field indicates the emergence of a highly unusual superconducting state.

In atomically thin films of CeCoIn$_5$,  the strong spin-orbit interaction in the presence of inversion symmetry breaking can dramatically affect the superconducting properties. The asymmetry of the potential in the direction perpendicular to the 2D plane $\Delta V\parallel$ [001] induces the Rashba spin-orbit interaction $\alpha_R g({\bm k})  \cdot {\bm \sigma} \propto ({\bm k} \times \Delta V) \cdot {\bm \sigma}$,  where $g({\bm k})=(k_y,k_x,0)/k_F$, $k_F$ is the Fermi wavenumber, and ${\bm \sigma}$  is the Pauli matrix.  The Rashba spin-orbit interaction splits the Fermi surface into two sheets with different spin structures. The energy splitting is given by $\alpha_R |k|$, and the spin direction is tilted into the plane, rotating clockwise on one sheet and anticlockwise on the other.  When the Rashba splitting exceeds the superconducting gap energy ( $\alpha_R |k|$$>\Delta$), the superconducting state is dramatically modified.   

The dramatic suppression of the Pauli effect in the trilayer CeCoIn$_5$ is naturally explained by this momentum-dependent splitting of spin bands. In the presence of an external magnetic field satisfying $\alpha_R |k|$$\gg \mu_B H$, the quasiparticle energy dispersion in the Rashba system is given as $E_{\pm}=\xi({\bm k})\pm\alpha_R|{\bm k}|\mp\mu_B g({\bm k})\cdot{ \mu_0\bm H}$, where $\xi({\bm k})$ is the quasiparticle energy without Rashba term and magnetic field and  $\mu_B$ is the Bohr magnetic moment. The anisotropic Zeeman interaction given by $\mu_B g({\bm k})\cdot{\bm H}$ leads to a strong suppression of the Pauli effect  for ${\bm H}\parallel [001]$ where $g({\bm k})\cdot{ \mu_0\bm H}=0$.

The enhancement of the orbital upper critical field in trilayer CeCoIn$_5$ is stunning, because it cannot be simply explained by the Rashba effect~\cite{kaur2005helical,samokhin2008upper}. We note that a strongly enhanced $H_{c2\perp}^{orb}$ was also observed in two other thin films, confirming that the enhancement arises from an intrinsic nature (see Fig.\,S9~\cite{SI}).  We note that in Fig.\,S9, the magnetic field does not significantly suppress the gap area. This is consistent with a theoretical calculation~\cite{nakai2006ubiquitous} showing that the gap size is similar to that at ${\bm H}$\,=\,0 even at ${ H/H_{C2\perp}}$\,=\,0.42.  Given the estimation of $\mu_0H_{c2\perp}^{orb}\approx 30$\,T, it is naturally expected that the gap size does not change up to 11\,T. One plausible explanation of such a location dependence is the effect of a vortex, where the magnetic field suppresses the superconducting state. Such a vortex effect may lead to underestimating the upper critical field, but not vice versa, so it does not change our claim of the enhanced $H_{c2}$.

This enhancement is not caused by disorder, because for unconventional pairing, both $T_c$ and $H_{c2\perp}$ are suppressed by disorder~\cite{bauer2006thermodynamic,yamashita2017fully}. We note that the situation is different from conventional superconductors such as Sn or MgB$_2$~\cite{bang2019characterization,gurevich2003very}, where the enhancement of H$_{c2}$ with a tiny reduction of $T_c$ is observed in thin films.  The thin films of the above conventional superconductors are granular and in the dirty limit, and the mean free path $\ell$ is much smaller than the coherence length $\xi_0$. By reducing the thickness, $\ell$ further decreases due to the dimensional effect (i.e., $\ell_c$ is limited by the film thickness), and the effective coherent length $\xi$ also decreases and $H_{c2}$ is enhanced. On the other hand, in $d$-wave superconductors, the mean free path is required to be much longer than the coherence length ($\ell$ $>>$ $\xi_0$) . In fact, according to Abrikosov-Gorkov theory, $T_c$ vanishes when $\ell$ becomes comparable to $\xi_0$~\cite{bauer2006thermodynamic,yamashita2017fully}. Therefore, the $H_{c2}$ enhancement cannot be explained by the reduction of $\ell$.   Comparing the present result with the recent report of highly enhanced $H_{c2}$ in thin films of Re~\cite{womack2021extreme} might be interesting. Also, since the substrate is a non-magnetic metal YbCoIn$_5$, this enhancement cannot be attributed to the substrate effect, while the reduction of $T_c$ may be partly due to the strain effect. 

This anomaly is highlightened by considering $H_{c2\perp}^{orb}/ \Delta^2$. Given $H_{c2\perp}^{orb}$ = $\Phi_0/2\pi\xi^2$ and $\Delta = \hbar v_F/\pi\xi$, where $v_F$ is the Fermi velocity and $\xi$ is the coherence length, we obtain $H_{c2\perp}^{orb}/ \Delta^2 \propto m^*/E_F$, where $m^*$ is the effective  mass.  Since $\Delta$ and $H_{c2\perp}^{orb} (0)$ of trilayer CeCoIn$_5$ is almost half and doubled compared with the bulk, respectively, we simply expect that $m^*/E_F$ of the trilayer is almost 8 time larger than that of the bulk.  As shown in Fig.\,2(a), the hybridization gap of atomically thin films is not largely different from that of the bulk.  Therefore,  $m^*$, which is determined by the hybridization and local $f$-electron interaction strength, is expected to be similar in atomically thin films and bulk. Then, there are two scenarios to explain the striking enhancement of $H_{c2\perp}^{orb}/ \Delta^2$.   One tempting scenario is that $E_F$ is significantly reduced in thin films.  In this case, $\Delta/E_F$ is largely enhanced and the system approaches the BCS-BEC crossover regime.  Another exotic  scenario is the modification of the pairing interaction in a magnetic field. In the absence of inversion symmetry, SOI gives rise to a parity-violated superconductivity. Such a superconducting state exhibits unique properties, including the admixture of spin-singlet and triplet states, which cannot be realized in superconductors with global inversion symmetry.  In fact, a microscopic calculation shows that a $p$-wave component can be induced that belongs to the same $B_1$ representation of the noncentrosymmetric $C_{4v}$ point group as the $d$-wave  and  these two components admix with each other~\cite{yoshida2016topological}. When the $p$-wave component is enhanced by the magnetic field, the orbital pair breaking effect may be weakened.

In summary,  {\it in-situ} STM study of atomically thin films of CeCoIn$_5$ reveal both hybridization and superconducting gaps even in monolayer CeCoIn$_5$, providing direct evidence of  heavy fermion superconductivity  mediated by purely 2D bosonic excitations.   Remarkably, the out-of-plane upper critical field  is largely enhanced from those of bulk and superlattices,  well exceeding the Pauli and bulk orbital limits.    This suggests the possible emergence of exotic superconductivity  caused by the broken inversion symmetry and clarifying its origin is a subject of future studies. 
 
We thank A. Daido, H. Kontani and Y. Yanase for fruitful discussions. This work is supported by Grants-in-Aid for Scientific Research (KAKENHI) (Nos. JP18H01180, JP18H05227, JP18K03511) from Japan Society for the Promotion of Science (JSPS), and JST CREST (JP-MJCR19T5).

\newpage

\end{document}


\title{Supplementary information of exotic heavy fermion superconductivity in atomically thin CeCoIn$_5$ films}
\author{L. Peng$^{1}$, M. Naritsuka$^{1,2}$,  S. Akutagawa$^{1}$,  S. Suetsugu$^{1}$,  M. Haze$^{3}$, Y. Kasahara$^{1}$, T. Terashima $^{1}$, R. Peters$^1$,  Y. Matsuda$^{1}$, and T. Asaba$^{1*}$}

\affiliation{
$^1$Department of Physics, Kyoto University, Kyoto 606-8502, Japan.\\
$^2$School of Physics and Astronomy, University of St Andrews, North Haugh, St Andrews, Fife KY16 9SS, UK.\\
$^3$Institute for Solid State Physics, University of Tokyo, Kashiwanoha 5-1-5, Kashiwa, Chiba 277-8581,  Japan.\\
}
\date{\today}
\maketitle

\newpage
\renewcommand\thefigure{S\arabic{figure}}

\noindent
 {\bf Formation of hybridization gap}
	
In the main text, we showed the asymmetric hybridization gap from atomically thin films of CeCoIn$_5$. The hybridization gap is asymmetric because of the gap formation process. Shown in Fig.\,S1 is the schematic diagram of the hybridization gap formation. Below the Kondo temperature, localized f-electrons hybridize with conduction electrons, opening the hybridization gap (Fig.\,S1). Since dI/dV is proportional to the local density of states, the gap becomes symmetric with respect to the Fermi energy only when (1) the DOS is symmetric to the gap center and (2) the Fermi energy locates at the gap center. Generally, neither condition is satisfied, resulting in an asymmetric hybridization gap.

\noindent
 {\bf Derivative of the spectra}
	
For the second derivative curves, we have also tried to first fit the data using Gaussian+polynomial, then take the second derivative of the fitted curves(Fig.\,S2). The particle-hole symmetry is also observed.

\begin{figure}[h]
	\centering
	\includegraphics[width=0.9\linewidth]{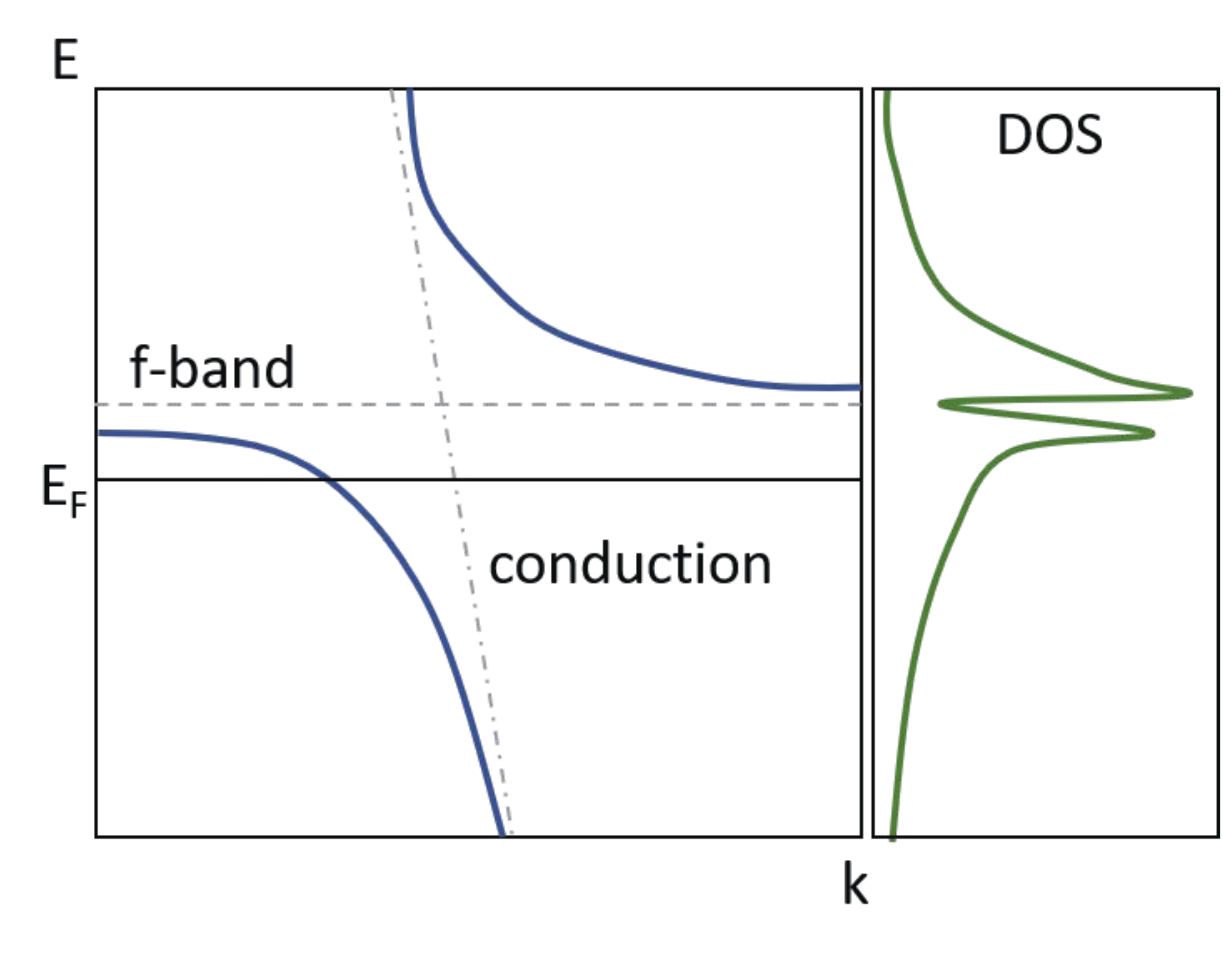}
	\caption{Schematic image of the formation of hybridization gap.}
\end{figure}

\begin{figure}[h]
	\centering
	\includegraphics[width=0.9\linewidth]{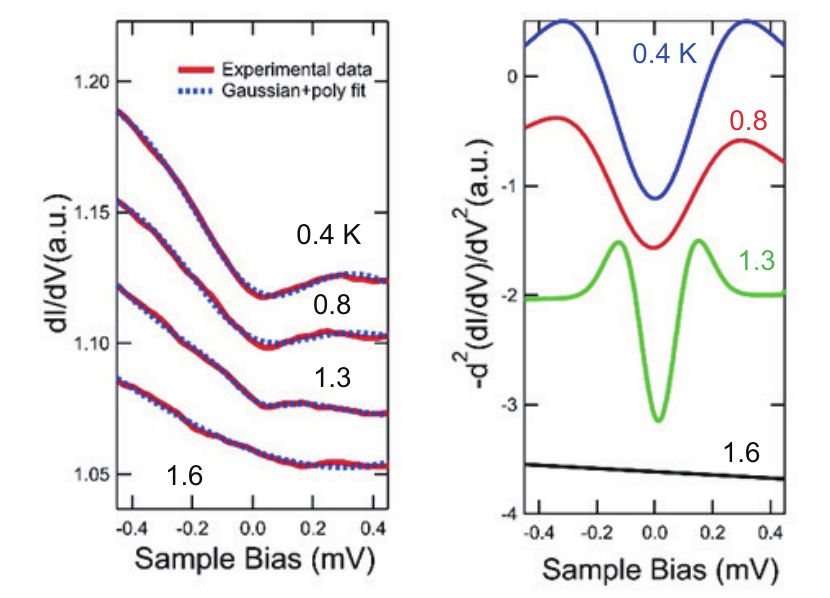}
	\caption{(left) The Gaussian+third-order polynomial fitting of the trilayer data. (right) The second derivative of fitted dI/dV curves.}
\end{figure}

\begin{figure}[h]
		\centering
	\includegraphics[width=0.9\linewidth]{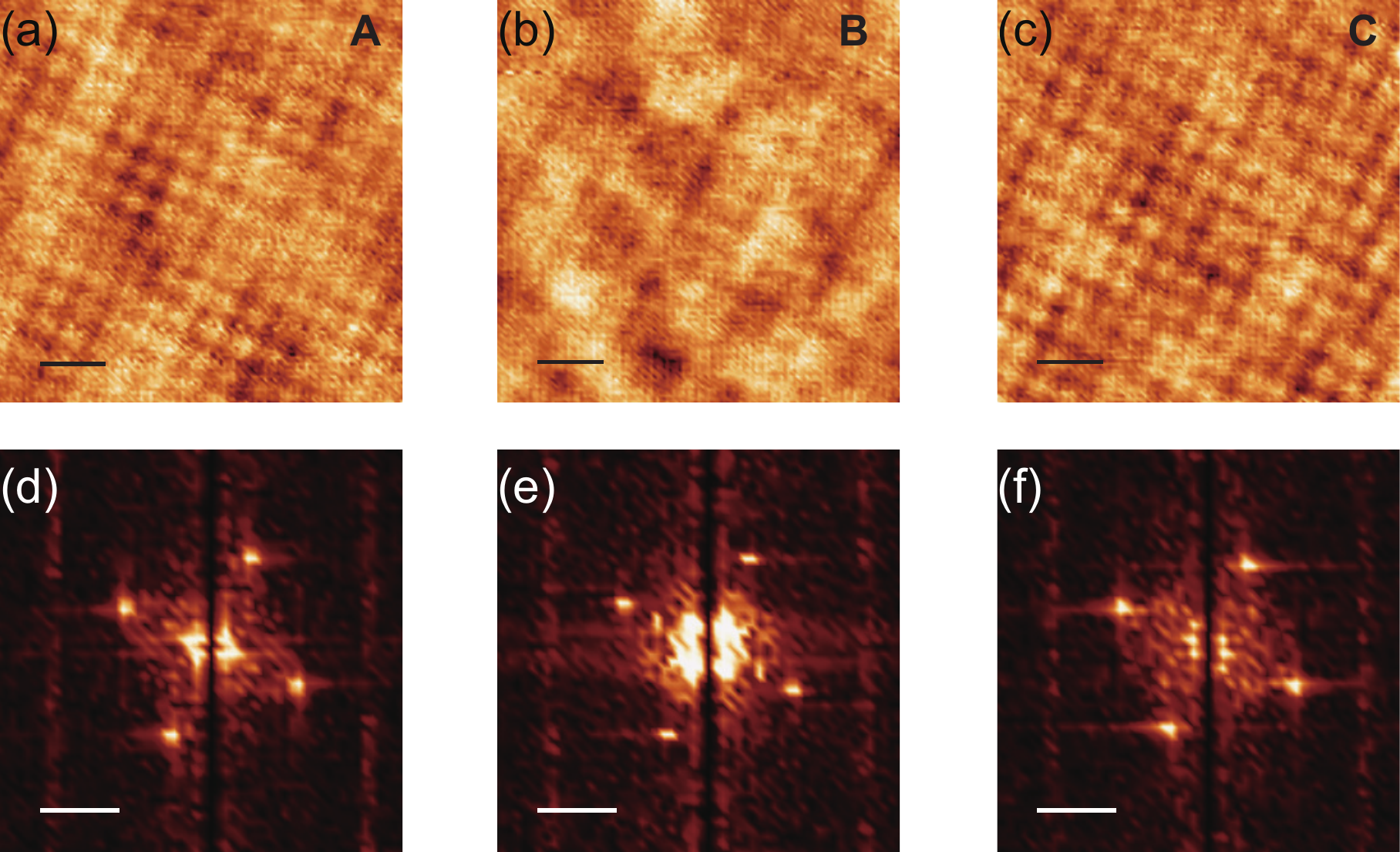}
	\caption{\label{SI_Fig:basic}  (a)-(c) STM images ($V_s = $100 mV, $I_t = $50 pA) taken on terrace A, B and C in Fig. 1(c), respectively.  For (a) and (c), bright spots that form the square lattice represent the In atoms in the Ce-In plane. For (b), bright spots represent the Co atoms.  Black bars denote $1$\,nm. (d)-(f) Fourier transform images of (a)-(c), respectively. Scale bar: 1.9\,nm$^{-1}$.}
\end{figure}

\begin{figure}[h]
		\centering
	\includegraphics[width=\linewidth]{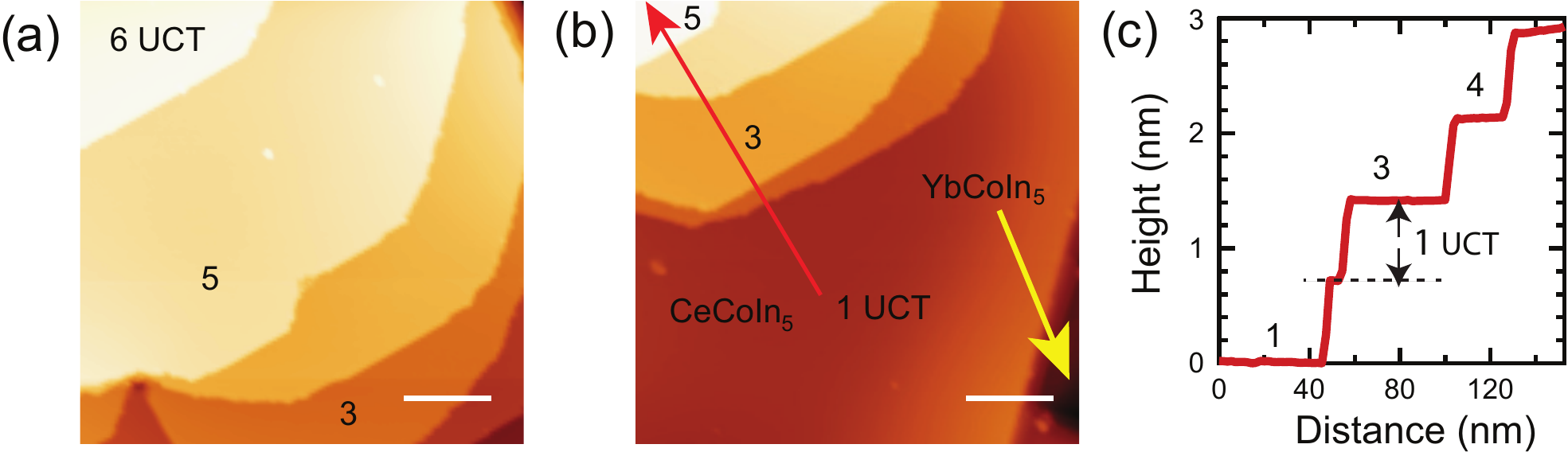}
	\caption{ (a)(b) Large scale STM images. The right-bottom corner of (a) is connected to the left-top corner of (b). White bars denote 40\,nm. (c) A line section along the red arrow in (b). Each step height is consistent with the $c$-axis lattice constant of CeCoIn$_5$. Image conditions: $V_s = $2.0 V, $I_t = $50 pA.}
\end{figure}

\begin{figure}[h]
		\centering
	\includegraphics[width=0.5\linewidth]{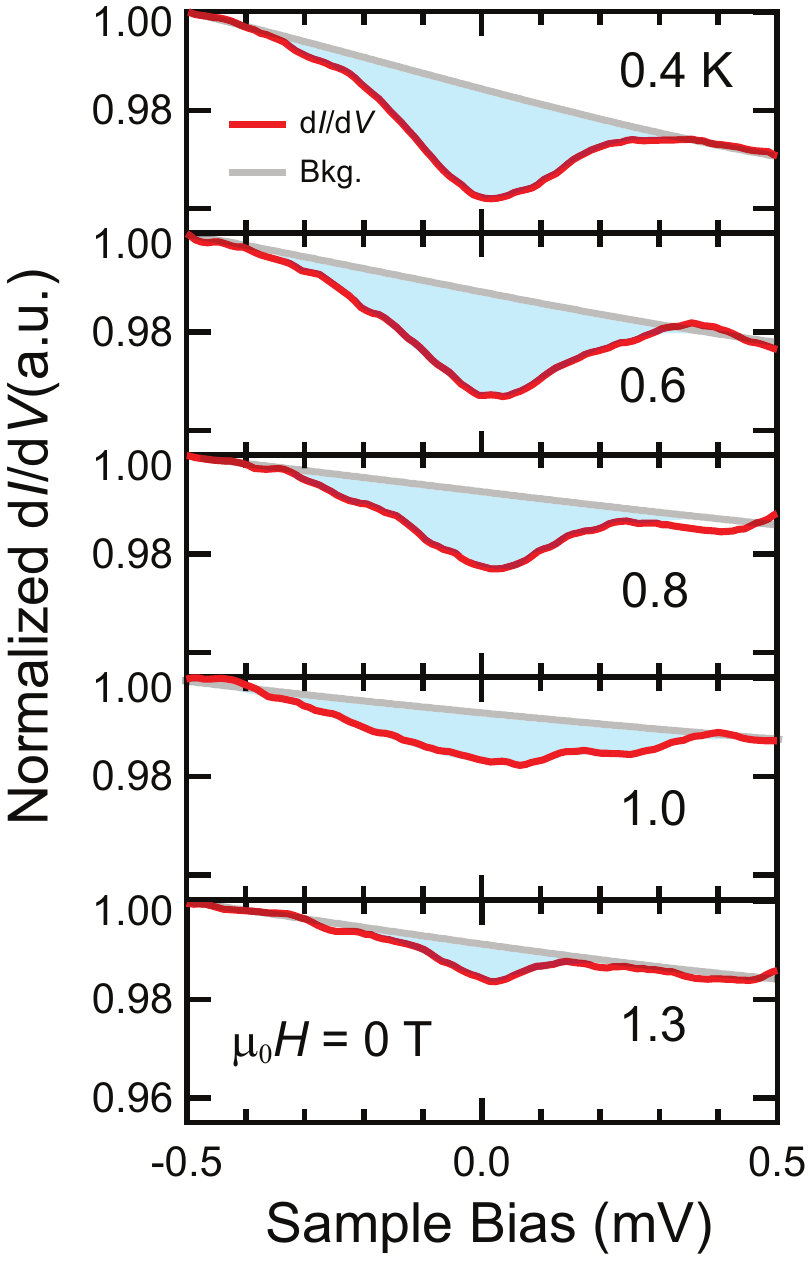}
	\caption{Temperature dependence of tunnelling conductance d$I$/d$V$ spectra in zero magnetic field. The d$I$/d$V$ spectra are normalized by that at 1.6\,K, then normalized by d$I$/d$V$ at V=-0.5\,mV.   Light blue region corresponds to superconducting gap area.  The gray lines represent the background obtained by cubic fitting in the high bias regime.}
\end{figure}

\begin{figure}[h]
		\centering
	\includegraphics[width=0.9\linewidth]{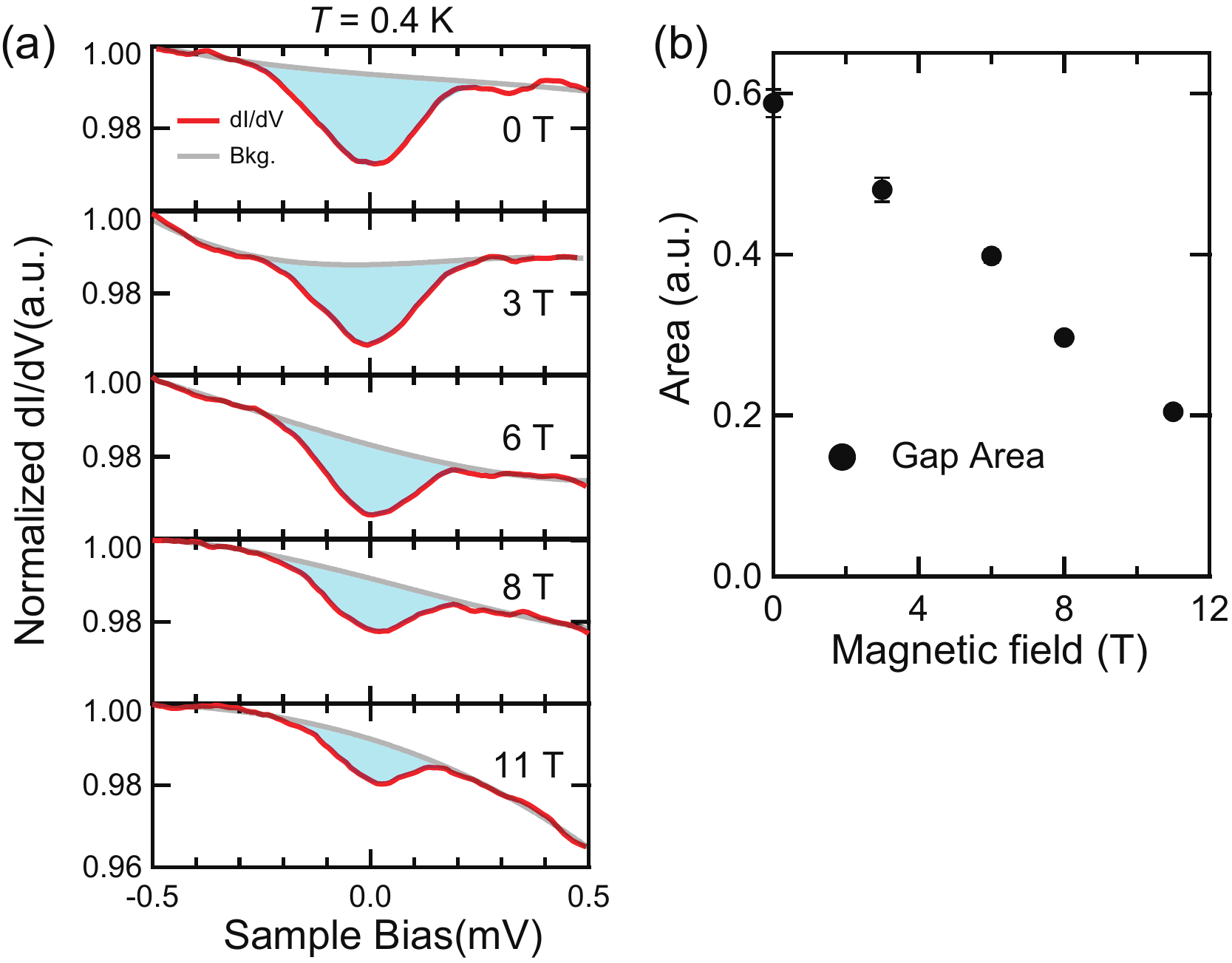}
	\caption{ (a) Field dependence of d$I$/d$V$ spectra at $T$=0.4\.K. The d$I$/d$V$ spectra are normalized by that at 1.6\,K, then normalized by d$I$/d$V$ at V=-0.5\,mV.   Light blue region corresponds to the superconducting gap area.  The gray lines represent the background obtained by polynomial fitting in the high bias regime. The SC gap still survives even at 11\,T. (b) Magnetic field dependence of the SC gap area. 
	}
\end{figure}

\begin{figure}[h]
		\centering
	\includegraphics[width=0.9\linewidth]{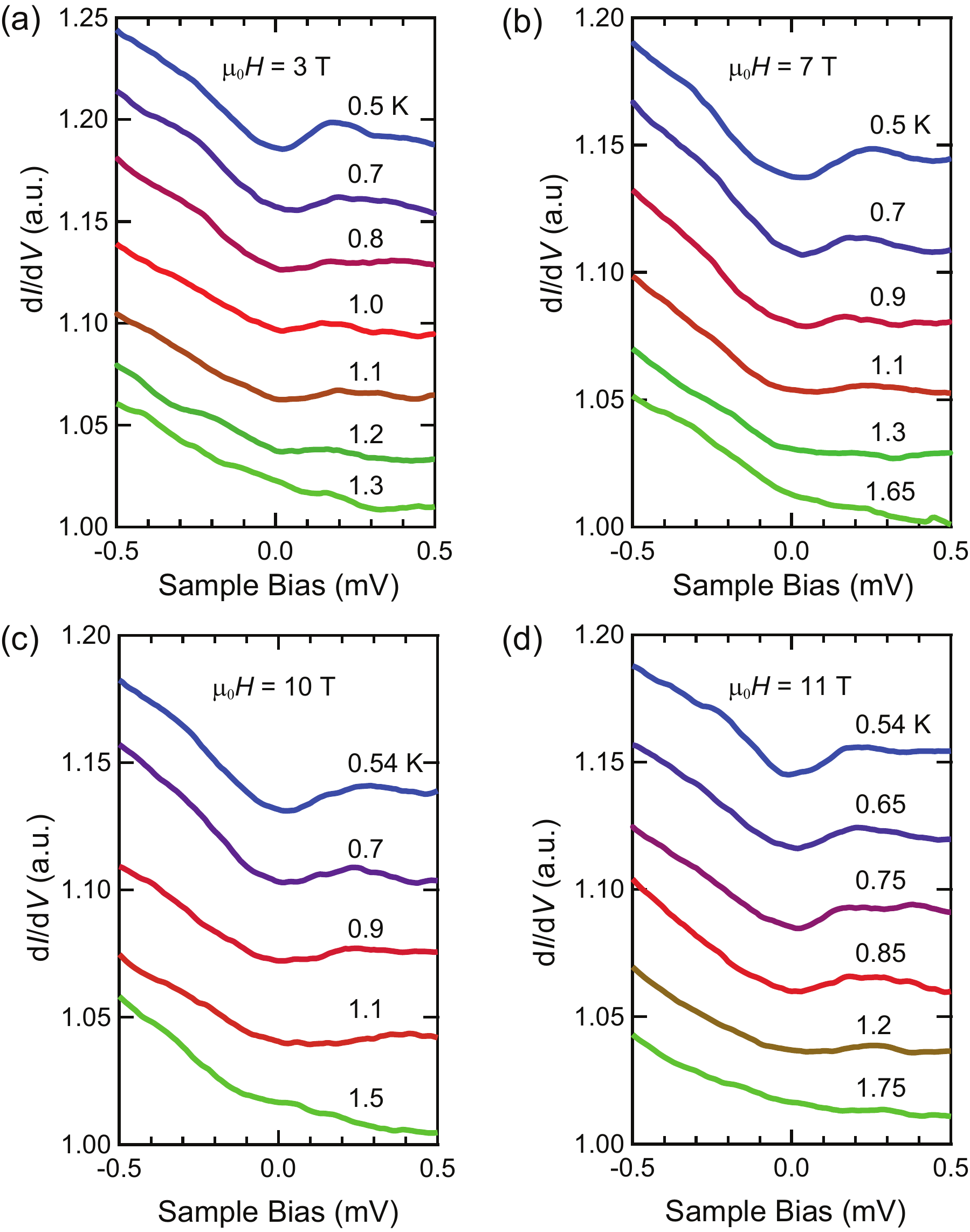}
	\caption{ (a)-(d) The applied field is $\mu_0 H $= 3, 7, 10 and 11\,T, respectively. Spectra are taken on the same site on the trilayer CeCoIn$_5$ surface. The spectra are vertically shifted for clarity.  Tunnelling parameters: $V_s=2$ mV, $I_t=50$ pA, $V_{mod}=30$$ \mu $V.}
\end{figure}

\begin{figure}[h]
		\centering
	\includegraphics[width=0.9\linewidth]{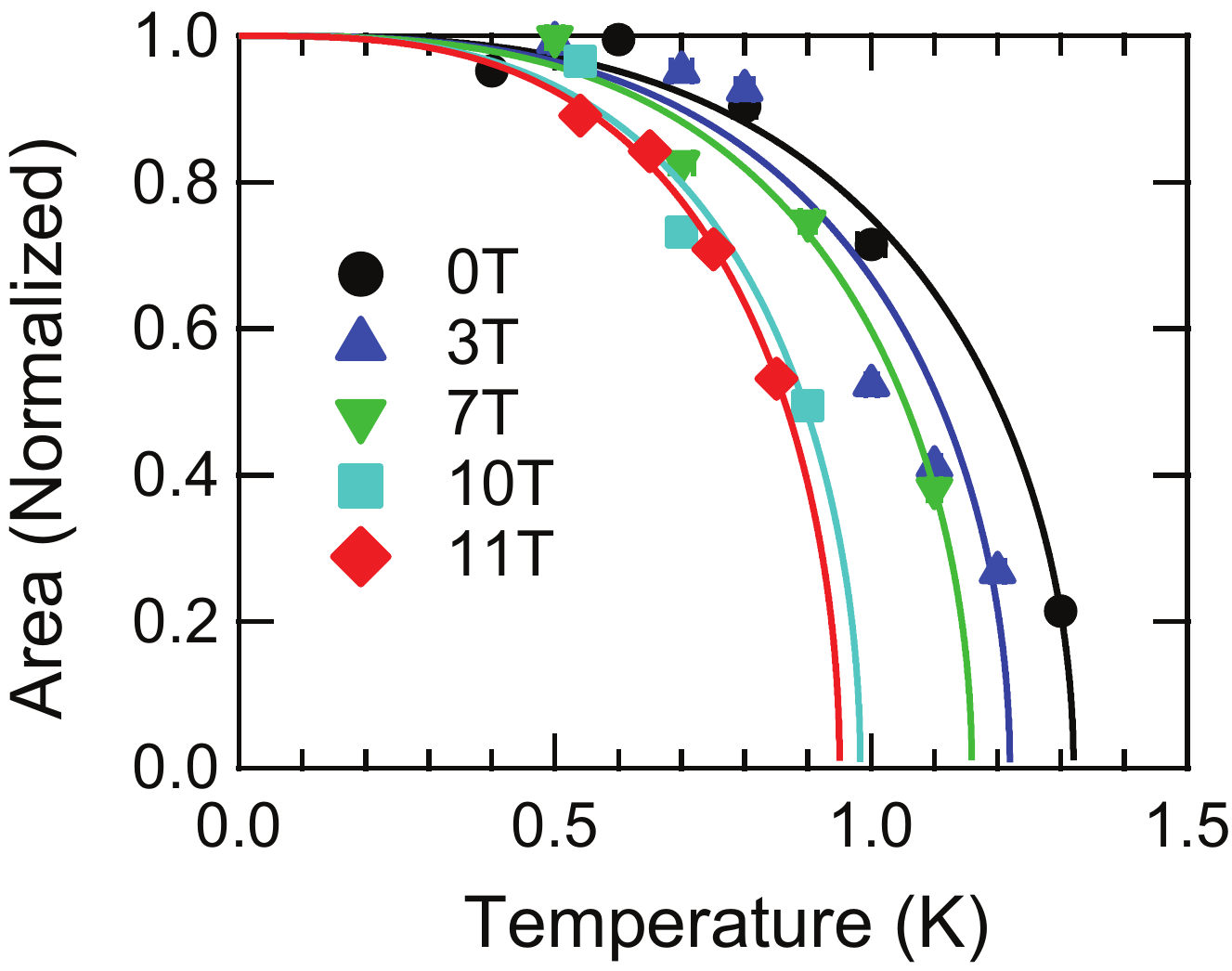}
	\caption{Temperature dependence of the superconducting gap area $S(T)$ at different magnetic fields. We fit the data by using $S(T) \propto \sqrt{1-(T/T_c)^3}$ at each field and obtain  $T_c$=1.22, 1.15, 0.98 and 0.95\,K at $\mu_0H$=3, 7, 10 and 11\,T, respectively. The data at zero temperature are normalized to 1 ($S (T\,=\,0)\,=\,1$). }
\end{figure}

\begin{figure}[h]
		\centering
	\includegraphics[width=0.9\linewidth]{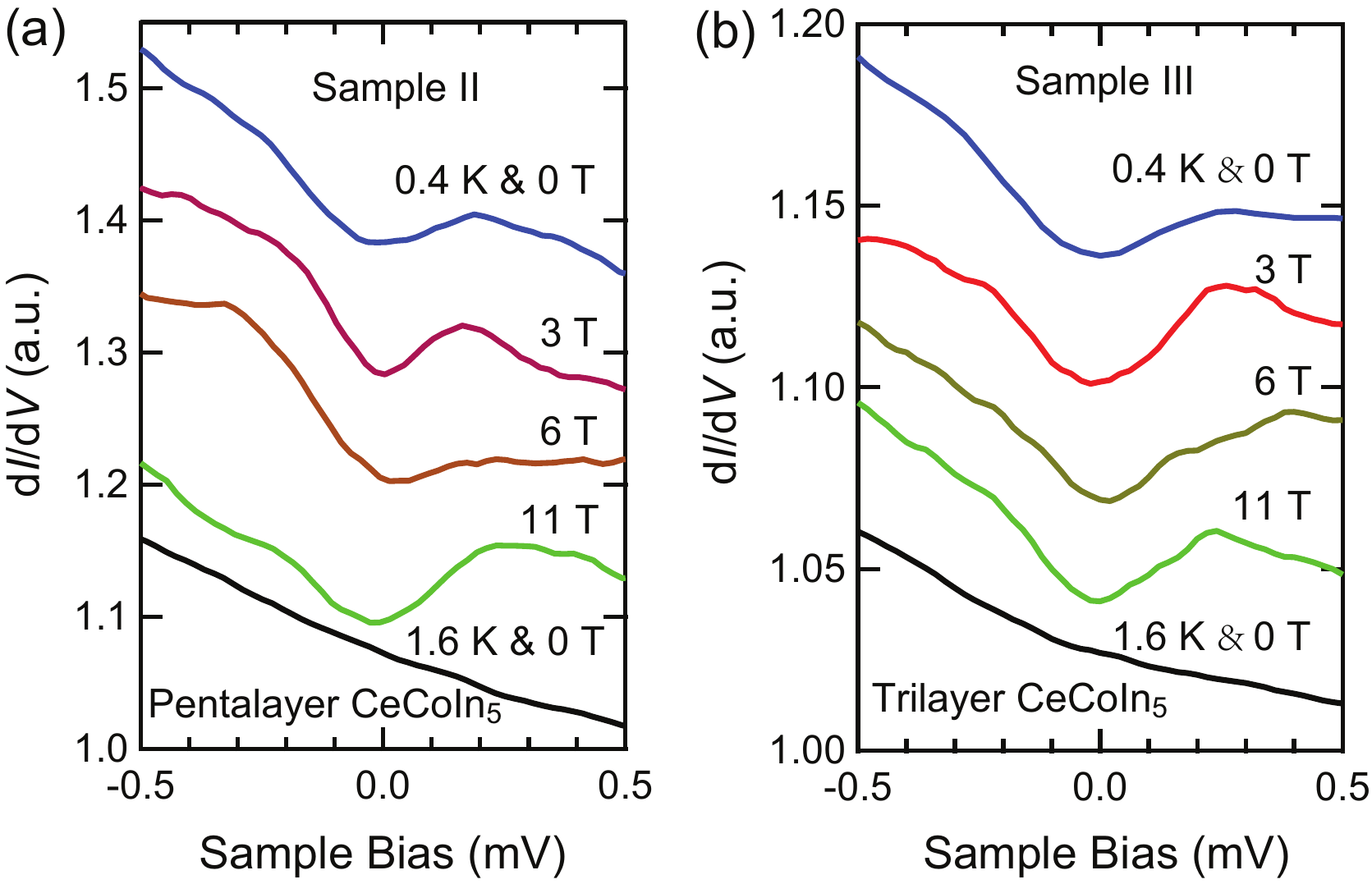}
	\caption{Robust superconductivity against magnetic fields in multiple samples. In addition to the sample shown in the main text, the survival of superconductivity at 11\,T was observed from two other samples.  As the position of the STM tip differs with different magnetic fields, the gap shape is influenced by inhomogeneity.
	}
\end{figure}

\begin{figure}[h]
		\centering
	\includegraphics[width=0.8\linewidth]{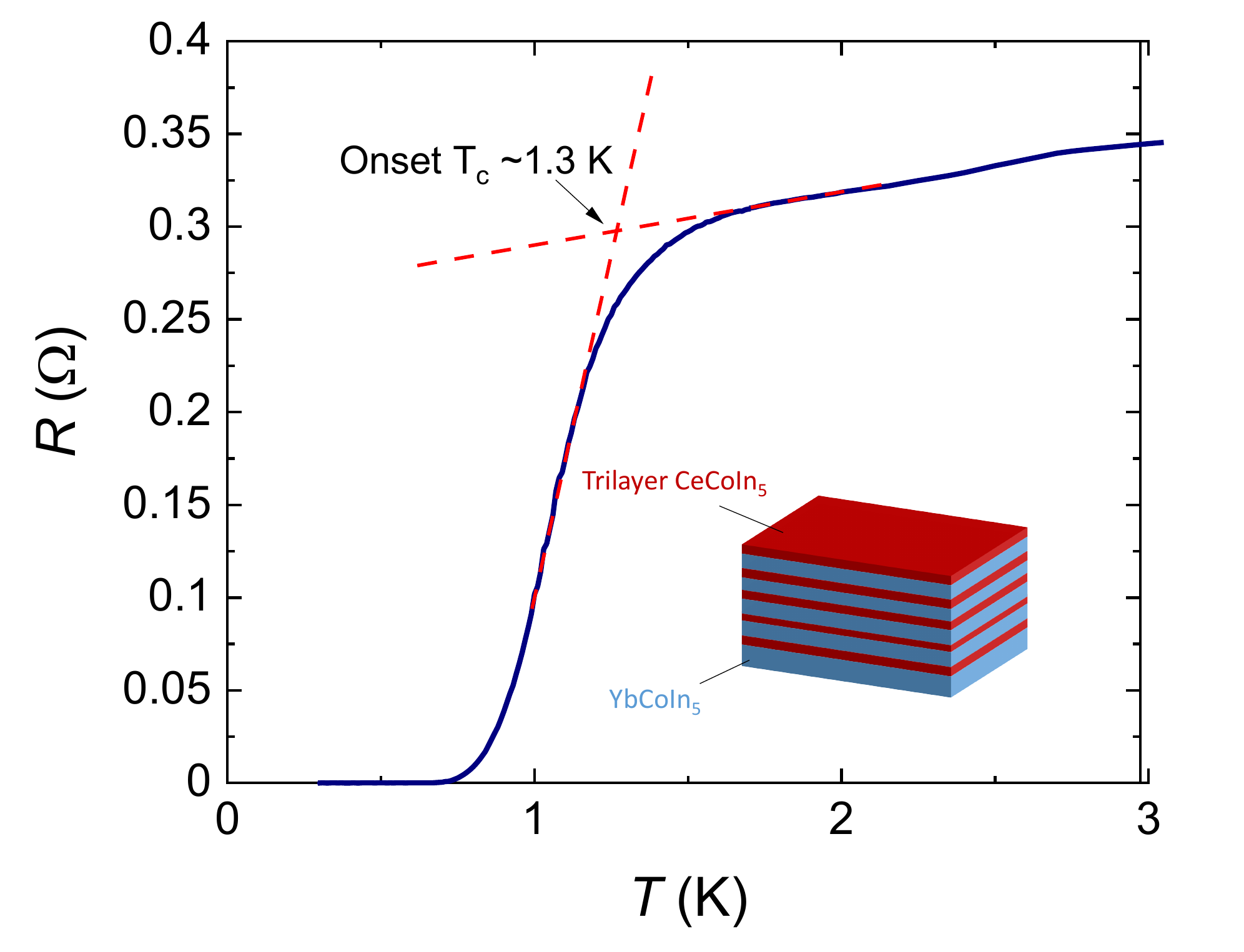}
	\caption{Resistivity curve of a trilayer CeCoIn$_5$ superlattice. Three layers of CeCoIn$_5$ are sandwitched by five layers of YbCoIn$_5$. The total thickness of the superlattice is $\sim$ 120\,nm. A clear superconducting transition with onset Tc$\sim$1.3\,K is observed.  
	}
\end{figure}

\begin{figure}[h]
	\centering
	\includegraphics[width=0.8\linewidth]{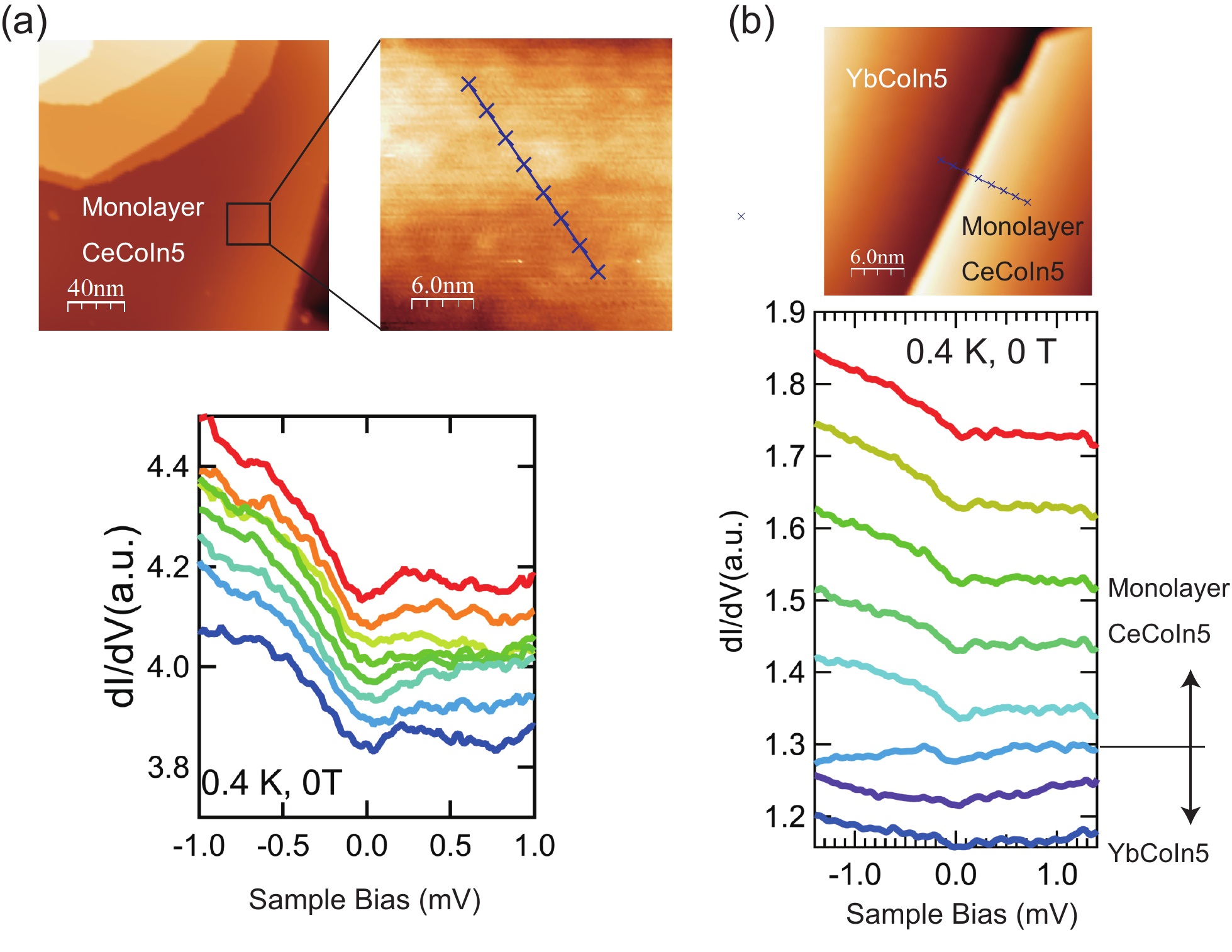}
	\caption{(a) dI/dV spectra of monolayer CeCoIn$_5$ taken along the line. (b) dI/dV spectra of the edge between YbCoIn$_5$ and CeCoIn$_5$ monolayer taken along the line.  The feature around zero bias on the bottom curve is either due to the experimental error or associated with the inhomogeneity of the sample.
	}
\end{figure}